\documentclass[12pt,final]{article}
\usepackage{amsmath,amssymb,amsfonts,amsthm,slashed,braket}
\usepackage[all]{xy}
\usepackage{mathrsfs}
\usepackage{fancyhdr}
\usepackage{showkeys}

\newcommand\nc{\newcommand}
\nc\linesep{\bigskip}
\nc\newprob[1]{\marginnote{#1}[\parskip]}
\nc\bA{\mathbb A}
\nc\bC{\mathbb C}
\nc\bD{\mathbb D}
\nc\bR{\mathbb R}
\nc\bZ{\mathbb Z}
\nc\bQ{\mathbb Q}
\nc\bP{\mathbb P}
\nc\bV{\mathbb V}
\nc\bW{\mathbb W}
\nc\bG{\mathbb G}
\nc\mf\mathfrak
\nc\mc\mathcal
\nc\mb\mathbb
\nc\brac[1]{\langle#1\rangle}
\nc\abs[1]{\lvert#1\rvert}
\nc\norm[1]{\lVert#1\rVert}
\nc\onto{\twoheadrightarrow}
\nc\into{\hookrightarrow}
\nc\lto{\longrightarrow}
\nc\action{\curvearrowright}
\nc\on\operatorname
\nc\wbar\overline
\nc\what\widehat
\nc\wtilde\widetilde
\nc\nop\DeclareMathOperator
\nc\eps{\varepsilon}
\nc\tsym{\widetilde{\text{Sym}}}
\nc\oarrow[1]{\overset{#1}\to}
\nop\Hom{Hom}
\nop\End{End}
\nop\Aut{Aut}
\nop\im{Im}
\nop\id{id}
\nop\tr{Tr}
\nop\coker{coker}
\nop\Spec{Spec}
\nop\Jac{Jac}
\nop\Ext{Ext}
\nop\Tor{Tor}
\nc\op{\text{op}}
\nop\loc{Loc}
\nop\Frac{Frac}
\nc\ann{\text{ann}}
\nop\QCoh{QCoh}
\nop\Coh{Coh}
\nop\Sym{Sym}
\nop\Hilb{Hilb}
\nop\gr{Gr}
\nop\Tot{Tot}
\nop\Fl{Fl}
\nop\tGamma{\widetilde\Gamma}
\nop\tloc{\widetilde{\text{Loc}}}
\nop\rep{Rep}
\nop\proj{Proj}
\nc\oo[1]{\overset\circ{#1}}
\nop\ospec{\oo{Spec}}
\nop\oTot{\oo{Tot}}
\nop\codim{codim}
\nop\holim{\underset{\lto}{holim}}
\nop\dlim{\underset{\lto}{lim}}

\nop\uHom{\underline{\Hom}}
\nop\dimrel{dim.rel}
\nc\sHom{\mathscr Hom}
\nc\sExt{\mathscr Ext}
\nc\dto{\dashrightarrow}
\nop\rspec{\bf Spec}
\nop\Gal{Gal}
\nop\Ind{Ind}
\nop\Frob{Frob}

\theoremstyle{theorem}

\theoremstyle{remark}

\begin{document} 

\centerline{\large{The hidden symmetry of the heterotic string}}
\bigskip
\bigskip
\centerline{Shamit Kachru$^1$ and Arnav Tripathy$^2$}
\bigskip
\bigskip
\centerline{$^1$Stanford Institute for Theoretical Physics}
\centerline{Stanford University, Palo Alto, CA 94305, USA}
\medskip
\centerline{$^2$Department of Mathematics, Harvard University}
\centerline{Cambridge, MA 02138, USA}
\bigskip
\bigskip
\begin{abstract}

We propose that Borcherds' Fake Monster Lie algebra is a broken symmetry of heterotic string theory compactified on $T^7 \times T^2$. As evidence, we study the fully flavored counting function for
BPS instantons contributing to a certain loop amplitude.  The result is controlled by $\Phi_{12}$, an automorphic form for $O(2, 26, \mb{Z})$. The degeneracies it encodes in its Fourier coefficients are graded dimensions of a second-quantized Fock space for this large symmetry algebra. This construction provides a concrete realization of Harvey and Moore's proposed relationship between Generalized Kac-Moody symmetries and supersymmetric string vacua.

\end{abstract}

\newpage

\tableofcontents



\section{Introduction}

It has long been a hope that string theory exhibits infinite-dimensional symmetries which constrain the form of string effective actions and S-matrices \cite{Gross, Giveon, Moore, HarveyMoore}.   Natural candidates are symmetries which generalize the normal structure of Lie groups and even affine Kac-Moody symmetries -- the so-called Generalized Kac-Moody (GKM)
algebras.
The Cartan matrices of these proposed larger symmetry algebras allow for negative diagonal entries, thereby greatly enlarging the root systems familiar from Lie theory to include whole lattices of ``imaginary roots''. That these extremely large symmetries might indeed exist as symmetries of sufficiently supersymmetric theories is of obvious interest for constraining the theory and possibly even formulating a classification of such theories. As such, the question has generated a considerable literature (including, in addition to the references above, papers such as \cite{HarveyMooretwo,Zwiebach,Gaberdiel,West} and references therein).

\medskip

Harvey and Moore provided a provocative new insight as to the source of these algebras in the context of 4d $\mc{N} = 2$ theories, arguing that the BPS states themselves have an algebraic structure
which can be inferred from projecting operator product expansions back to the BPS sector \cite{HarveyMoore, HarveyMooretwo}.   
They argued that in simple cases, this algebraic structure indeed matches up well with the structure of Generalized Kac-Moody algebras.\footnote{As per the usage of~\cite{HarveyMooretwo}, we mean something slightly more general by a GKM than Borcherds' original definition due to allowing more general gradings than simply by $\mb{Z}$.}  
Much of the recent work has focused attention on the 4d $\mc{N} = 4$ case as more amenable to progress.  In such theories, we have the ability to encapsulate the full BPS-counting functions and can thereby test that an appropriate particle spectrum does organize into a representation of the symmetry (c.f. \cite{DVV,ChengVerlinde,Sureshone,Sureshtwo,Sureshthree,Sureshfour}).  Most recently, Paquette-Persson-Volpato argue for a Monstrous Lie algebra symmetry in an asymmetric orbifold of the heterotic string which they studed for moonshine purposes \cite{PPV1, PPV2}. 

\medskip

Here, we propose a new entrant to this list of examples, providing both a concrete proposed Generalized Kac-Moody algebra and an `instanton counting' generating function that may be clearly written as a character of a specified representation. The instanton counting function roughly arises from a one-loop amplitude in heterotic string theory, and can be computed on $T^7 \times T^2$ with a large $T^2$.  (In defense of our compactification to 0+1 dimensions, we note in passing that this compactification is a relatively conservative setting in this particular subject \cite{Moore}.)  In fact, we aim to go further: we explain why the Weyl denominator formulas discovered in previous studies of GKMs in string theory are exactly in line with the Harvey-Moore predictions. Indeed, we argue that the full instanton spectrum organizes into a second quantized Fock space on the creation operators of the symmetry algebra. This space is endowed with the structure of a $\mf{g}$-representation as a Verma module for the algebra.\footnote{It would be more conventional to talk about a spectrum of BPS particles, but in low dimensions this notion is problematic; charged and gravitating particles have large backreaction.  Instead, we simply discuss the spectrum of states visible in formulae for BPS-protected terms in the 3D effective action \cite{Borisone,Boristwo}.  Because in various asymptotic limits of moduli the states running in these formulae have interpretations as higher dimensional BPS states propagating around a circle, we call the resulting function an instanton count.}

\medskip

Concretely, we propose that Borcherds' Fake Monster Lie algebra, as explicated in \cite{Borcherds, Borcherdstwo}, is a broken symmetry of heterotic string theory compactified on $T^7 \times T^2$.  
The spectrum of BPS states running in a suitable loop amplitude organizes into a representation of this algebra as above.  The Fake Monster has rank 26, and the 26 gradings (eigenvalues of the Cartan generators) are
charges of the instantons, governing their action.  (Colloquially, they would govern the masses of the related particles.)   The full set of states is 
organized into a simple highest-weight representation of the symmetry. Moreover, this function is the inverse of the celebrated Borcherds automorphic form $\Phi_{12}$, following \cite{five}; this function is automorphic for  $O(2, 26;{\mathbb Z})$.  

\medskip
It is our hope that the appearance of such a large symmetry group may eventually help in finding an exact solution of (a suitable sub-sector of) this string theory.

\medskip

We begin in the next section by reviewing past work on GKM algebras for half-maximally supersymmetric theories.  In section $3$ we recall results on the $1/2$-BPS instanton spectrum of the heterotic string on $T^7
(\times T^2) $. In sections $4$ and $5$, we give general remarks on the representation-theoretic content of the inverse Weyl denominator formula for GKMs, arguing for a general understanding of these automorphic forms as following from the Harvey-Moore prescription for BPS algebras.  Section $4$ recalls basic results on the Weyl character formula before section $5$ explains the particular representation-theoretic import of the inverse Weyl denominator.  Section $6$ then applies this logic to argue for the Fake Monster Lie algebra as a symmetry of the toroidally compactified heterotic string.  We conclude in section $7$. Finally, in an appendix, we discuss the relation of our representation-theoretic constructions to that of similar work in the mathematical and physical literature \cite{Kostant, Kostanttwo, PPV1, PPV2} using instead a fermionic Fock space construction. 


\section{Review: GKMs in theories with half-maximal SUSY}

The well-known examples of string vacua with sixteen real supercharges are the toroidally compactified heterotic string or the type II string on $K3$ and some number of circles, possibly with extra orbifolding to give rise to the CHL models.   Here we will just focus on the simplest (and original) example.

\medskip
One of the first concrete relations between GKMs and string vacua to be proposed involved the 4d $\mc{N} = 4$ theory arising in type II strings on $K3 \times T^2$. 
The GKM naturally emerged in attempts to understand counting functions for BPS degeneracies, with the ultimate goal of giving microscopic explanations of black hole entropy \cite{DVV}.
This theory admits both $1/2-$ and $1/4-$BPS particles, the former with only one U-duality invariant (electric charge squared, say), and the latter with three $U$-duality invariants of $Q_e \cdot Q_e, Q_e \cdot Q_m,$ and $Q_m \cdot Q_m$, where $Q_e, Q_m$ represent the electric and magnetic charges, respectively. The one-variable $1/2$-BPS state generating function is easily calculated in the dual heterotic frame to be $$\sum c_n q^{n-1} = \frac{1}{\eta(\tau)^{24}} = \frac{1}{\Delta(\tau)},$$ where $c_n$ is the count of states with charge squaring to $2n - 2$. The $1/4$-BPS state count is slightly more involved and was first explicated in \cite{DVV} to be $$\sum c_{n, \ell, m} p^{n} y^{\ell} q^{m} = \frac{1}{\Phi_{10}(\sigma, z, \tau)}.$$ Here, $c_{n, \ell, m}$ the number of multiplets with charges satisfying $$\frac{1}{2} Q_e \cdot Q_e = n,~~ Q_e \cdot Q_m = \ell, ~~\frac{1}{2} Q_m \cdot Q_m = m,$$ and $\Phi_{10}(\sigma, z, \tau)$ is the (unique) Siegel cusp form of degree $2$, weight $10$, and full level. 

\medskip
We are interested here in the organization of particle states into representations of symmetry algebras so that the above generating functions organize as characters. That the $1/4$-BPS generating function relates to a large symmetry algebra is already an interesting speculation in \cite{DVV} itself, where the authors note that that $\Phi_{10}$ is the Weyl-Kac denominator of a rank $3$ Generalized Kac-Moody algebra found in \cite{GritsenkoNikulin} with Cartan matrix $$ \begin{pmatrix} -2 && 2 && 2 \\ 2 && -2 && 2 \\ 2 && 2 && -2 \end{pmatrix}.$$  As an aside, we note that the simpler question of the symmetry underlying the $1/2$-BPS generating function is arguably answered by the Heisenberg algebra construction of Nakajima \cite{Nakajima} in his proposed categorification of the Vafa-Witten invariants of $K3$ \cite{VafaWitten}.  A careful explanation of how this statement embeds in the more intricate stories such as $1/4$-BPS counts or more flavored counts has yet to be provided either mathematically or physically.


\medskip
Our work will extend these works both by going to lower dimensions, but more importantly, by considering fully flavored counting functions.  Instead of using U-duality invariance to restrict to a small number of invariants, as in \cite{DVV}, we will flavor by the full set of electric charges available in the theory.  This gives rise to a function with 26 gradings, which we are able to match to the denominator of the Fake Monster.  We further specify the representation of the Fake Monster whose character gives the BPS counting function we identify.

\medskip

Earlier work that is similar to ours in considering higher rank GKMs was performed by Gaberdiel, Hohenegger, and Persson \cite{Gaberdiel}.
The authors consider BPS states in heterotic string theory on $T^2$.  
They discuss the 1/2-BPS Dabholkar-Harvey states,
and obtain a rank $18$ GKM by considering the full set of gradings arising from left-moving momentum quantum numbers, where the rank is $18$ as the relevant Narain moduli space in 8d is $O(2,18,{\mathbb Z}) \backslash O(2,18) / (O(2) \times O(18))$.  These authors also discuss a natural rank $26$ extension of their structure (very roughly arising by adjoining another $E_8$ lattice to their construction).  
We believe their story is a natural relative of ours, and that the Fake Monster Lie algebra is the hidden symmetry structure.



\section{The heterotic string on $T^7 (\times T^2)$}

The Narain moduli space of the heterotic string on $T^7$ is
$$O(7,23, \mb{Z}) \backslash O(7,23) / (O(7) \times O(23))~.$$
However, using the duality of vectors to scalars in three dimensions as well as the existence of a large
group of U-duality symmetries, Sen \cite{Sen} determined that the actual moduli space enhances to the double-coset
 $${\cal M}_{3D} = O(8, 24, \mb{Z}) \backslash O(8, 24) / (O(8) \times O(24))~.$$ 

\medskip
This can be thought of as the moduli space of even self-dual lattices of signature $(8,24)$.  It is convenient to choose a 
a point in ${\cal M}_{3D}$ where the lattice splits as
$$\Gamma^{8,24} = E_8 \oplus N(-1)$$
where $E_8$ is the $E_8$ root lattice, and $N$ is one of the 24 even self-dual lattices in dimension 24 (the 23 Niemeier lattices, and the Leech lattice).  At such a 
point, as discussed in e.g. \cite{five}, the graded counting function for 1/2-BPS states can be written as 
$${\Theta_{N}(\tau,\xi) \over \eta^{24}(\tau)}$$
where
$$\Theta_{N}(\tau,\xi) \equiv \sum_{\lambda \in N} e^{\pi i \tau (\lambda,\lambda) + 2\pi i (\xi,\lambda)}$$
with $(\cdot,\cdot)$ the inner product on $N$.  Here, $\xi$ should be thought of as a 24-vector of (complex) chemical potentials grading the BPS states under
the 24 left-moving abelian currents of the heterotic string on $E_8 \oplus N(-1)$.

\medskip

Further compactifying on a $T^2$ has the following effect.  A priori, the moduli space would enlarge to (at least) the Narain moduli space based on even, self-dual signature (10,26)
lattices.  However, we are interested only in considering a product compactification $T^7 \times T^2$ where we permit Wilson lines of the 24 gauge fields in matter multiplets of the 3D theory, but not metric or B-field mixing.  The resulting moduli space parametrizes lattices of signature $(2,26)$, since we have kept an $E_8$ factor isolated
on the right. 

\medskip

Now, the BPS instanton count will be given by the multiplicative lift of the 3D BPS counting function, as in \cite{DMVV}.  The result, as discussed in \cite{five}, is
the inverse of the Borcherds modular form $\Phi_{12}$.\footnote{For foundational expositions of its properties, see e.g. \cite{BorcherdsExp} and \cite{Gritsenko}.}  This is a form on
$$O(2,26, \mb{Z}) \backslash O(2,26) /(O(2)\times O(26))~.$$
For a given choice of $N$, we obtain this form expanded around a ``Niemeier cusp'' in this moduli space.
Let us define the Fourier coefficients of the 3D 1/2-BPS instanton counting function as 
$${\Theta_N(\tau,\xi) \over \eta^{24}(\tau)} = \sum_{n \in {\mathbb Z}, \lambda \in N} f(n,\lambda) q^n e^{2\pi i (\xi,\lambda)}~.$$
Then $\Phi_{12}$, expanded around the given Niemeier cusp, is
$$\Phi_{12}(\tau,\xi,\sigma) = q^A r^{\vec{B}} p^C \prod_{\substack{n,m \in {\mathbb Z}\\ \lambda \in N \\ (n, l, m)>0}} ~(1-q^n r^\lambda p^m)^{f(mn,\lambda)}$$
$$p \equiv e^{2\pi i \sigma}, ~r^\lambda = e^{2\pi i(\xi,\lambda)}.$$
Here, 
$$A = {1\over 24} \sum_{\lambda \in N} f(0,\lambda),~~{\vec B} = {1\over 2}\sum_{\lambda > 0} f(0,\lambda) \lambda,~~C = {1 \over 48} \sum_{\lambda \in N} 
f(0,\lambda) (\lambda,\lambda)^2~.$$
The notation $(n,l,m)> 0$ in the formula for $\Phi_{12}$ means $m, n \ge 0$, with $\lambda < 0$ if $m = n = 0$. Here, $\lambda > 0$
(or $\lambda < 0$) means that $\lambda$ has a positive (negative) inner product with some reference vector $x \in N \otimes {\mathbb R}$.  The vector $x$ should
have $(x,\lambda) \neq 0$ for all $\lambda \in N$.

\medskip

In the sequel, we will show that $\Phi_{12}$, which we are calling a BPS counting function with the caveats explained previously, arises as a character of the Fake Monster Lie algebra, and that the states being counted by $\Phi_{12}$ furnish a representation.  This leaves the question of why we have considered only $T^7 \times T^2$ and not more generic geometries for the $T^9$.  Partly, this is to make contact with a known symmetry structure -- the Fake Monster.
But it has also been, at least in part, due to our lack of knowledge about automorphic forms for
$O(10,26;{\mathbb Z})$ (or perhaps the even larger U-duality group).  It would be very interesting to find a larger algebraic structure and a representation such that the character is automorphic for this group, and
the coefficients count BPS instantons in heterotic string theory on a generic $T^9$.

\section{The Weyl character formula}

We now explain some foundational material on the Weyl character and Weyl denominator formulas.  We present the ensuing discussion in the familiar case of finite Lie groups for the benefit of the moderately mathematically sophisticated reader. Similar statements and proof techniques hold for the case of affine Kac-Moody algebras, and while we wish to make general statements for the Generalized Kac-Moody symmetry algebras that arise as symmetries of supersymmetric string vacua, such general proofs (even of the existence of a Weyl vector!) are not known in the mathematics literature; we gloss over these concerns here.\footnote{In fact, many GKMs are known not to admit Weyl vectors~\cite{Ray}, but we consider this failure a looseness in the definition of GKMs. Certainly, for example, we would wish to only consider the subclass of GKMs whose Weyl denominators are automorphic forms.}

\medskip
We recall that for a finite-dimensional group $G$, representations $V_{\lambda}$ are labeled by their highest weight $\lambda$ and that weights of the adjoint are the specially denoted roots $\Delta$. In order to make sense of highest weights, we make a choice of positive roots $\Delta^+$ and let the Weyl vector $\rho$ be half the sum of positive roots. Then the Weyl character formula states the following: $$\mathrm{char}(V_{\lambda}) = \frac{\sum_{w \in W} \varepsilon(w) e^{w(\lambda + \rho)}}{e^{\rho} \prod_{\alpha \in \Delta^+} (1 - e^{-\alpha})}.$$ Here $W$ denotes the Weyl group and we interpret characters as Weyl-invariant functions on a maximal torus $T$. To finish explaining the meaning of the above formula, for $\lambda$ a weight, we denote $e^{\lambda}$ as the corresponding character $T \to \mb{C}^*$ so that the above expression is indeed a function on $T$. In fact, any Weyl-invariant function on the torus is the character of some (possibly reducible virtual) representation, and we may indeed verify that the above expression is Weyl-invariant -- both the numerator and denominator are explicitly anti-invariant under the Weyl action, i.e. acting by $w \in W$ yields the sign $\varepsilon(w)$ corresponding to the determinant of the orthogonal action of $w$ on the root space.

\medskip
Hence, for example, to find the dimension of the representation $V_{\lambda}$, we would evaluate its character at the identity element and evaluate all ``exponentials'' above to be $1$. Of course, as the Weyl character formula is written above, both numerator and denominator vanish to high order and after l'H\^{o}pital-style simplification, the dimension is given by the Weyl dimension formula $$\dim V_{\lambda} = \prod_{\alpha \in \Delta^+} \frac{\langle \lambda + \rho, \alpha \rangle}{\langle \rho, \alpha \rangle}.$$ 

\section{The inverse Weyl denominator}

The denominator of the Weyl character formula $$\Phi = e^{\rho} \prod_{\alpha \in \Delta^+} (1 - e^{-\alpha})$$ has played an especial role in the string theory literature on Generalized Kac-Moody algebras. We seek to explain how to interpret the representation-theoretic meaning of a BPS spectrum organizing itself into the inverse of such a Weyl denominator formula. Our claim is that an inverse Weyl denominator formula indicates that the particles are organized into a second-quantized Fock space built on the creation operators of $\mf{g}$. Indeed, most directly, we simply note that the inverse Weyl denominator $$\Phi^{-1} = \frac{e^{\rho}}{\prod_{\alpha \in \Delta^+} (1 - e^{-\alpha})} = \mathrm{char}\,\,M_{\rho},$$ where $M_{\rho}$ is the Verma module of highest weight the Weyl vector $\rho$, as per Section 4 of~\cite{BorcherdsChar}. We recall in the following the definition of a Verma module and why this construction is physically natural from the perspective of Harvey and Moore.

\medskip
First, we recall that the choice of positive roots at the outset provides a natural decomposition $$\mf{g} \simeq \mf{n}^+ \oplus \mf{h} \oplus \mf{n}^-.$$ Here $\mf{h}$ is the Cartan subalgebra, the Lie algebra of a maximal torus, while $\mf{n}^+, \mf{n}^-$ are the sums of the positive and negative root-spaces respectively. We also denote $$\mf{b} \simeq \mf{n}^+ \oplus \mf{h}$$ as the Lie algebra of the Borel. For example, for $\mf{g} \simeq \mf{gl}_n$ the Lie algebra of $n \times n$ matrices, the subalgebra $\mf{b}$ is the space of upper-triangular matrices while $\mf{n}^+$ ($\mf{n}^-$) is the space of strictly upper-triangular (lower-triangular) matrices. Per convention, we consider $\mf{n}^-$ as creation operators, $\mf{n}^+$ as annihilation operators, and the Cartan $\mf{h}$ the abelian algebra tracking the flavors. 

\medskip 
The universal enveloping algebra $U(\mf{g})$ of a Lie algebra $\mf{g}$ is the canonical associative algebra ``built on'' the Lie algebra $\mf{g}$. More precisely, it is the tensor algebra of $\mf{g}$, or the collection of free words on a basis for $\mf{g}$, quotiented by the commutativity relations defining the Lie algebra structure on $\mf{g}$. By choosing some lexicographic ordering on the basis elements and using the commutation relations to re-order all words per this lexicographic ordering, it is easy to see that the underlying vector space is isomorphic to that of the symmetric algebra $\Sym \mf{g}$. The Verma module $M_{\lambda}$ is now given by the tensor product $$M_{\lambda} = U(\mf{g}) \otimes_{U(\mf{b})} \mb{C}_{\lambda},$$ where $\mb{C}_{\lambda}$ is a one-dimensional space acted upon by the Borel through its quotient Cartan via the character $\lambda$. The particular determination of $\lambda$ just contributes a prefactor to the character and we ignore it further in this discussion, seeking to simply explain the underlying vector space; as such, we can set $\lambda = 0$ and consider the representation $U(\mf{g}) / (U(\mf{b}))$, where we quotient by the left-ideal generated by the Borel, i.e. set to zero any words ending with elements of the Cartan $\mf{h}$ or the annihilation operators $\mf{n}^+$. 

\medskip
For example, for $\mf{g} = \mf{su}(2)_{\mb{C}}$, the universal enveloping algebra $U(\mf{g})$ is the vector space on free words on $e, h,$ and $f$ subject to the usual commutation relations $ef - fe = h, he - eh = e, hf - fh = -f$. Quotienting by the left ideal $(U(\mf{b}))$ sets all words ending with $h$ or $e$ equal to zero, so that what remains is free words on the basis element $f$. In general, what remains are words on the creation operators, and so our construction is endowing the bosonic Fock space $\Sym \mf{n}^-$ with a natural structure as a $\mf{g}$-representation.

\medskip
We further explicate the example of $\mf{su}(2)_{\mb{C}}$. Here, the Weyl denominator is  $$\Phi = \prod_{\alpha \in \Delta^+} (1 - e^{-\alpha}) = 1 - t^{-1}$$ as there is simply one negative root corresponding to the annihilation operator $f$, on whom $h$ acts with eigenvalue $-1$. For these conventions, we wish to expand in powers of $t^{-1}$, and so we write the inverse Weyl denominator as $$\Phi^{-1} = \frac{1}{1 - t^{-1}} = 1 + t^{-1} + t^{-2} + t^{-3} + \cdots.$$ Let us now consider our proposed representation $$\Sym \mf{n}^- \simeq U(\mf{g}) / (U(\mf{b})) \simeq ( \mb{C} \cdot 1 ) \oplus ( \mb{C} \cdot f ) \oplus ( \mb{C} \cdot f^2 ) \oplus \cdots.$$ Recall that in the universal enveloping algebra, we have $hf = fh - f$; hence, after quotienting by $(U(\mf{b}))$ so that all words ending with $e$ or $h$ vanish, we have $hf = -f$. Similarly, we have $h \cdot 1 = h = 0$, $hf^2 = (fh - f)f = fhf - f^2 = f(fh - f) - f^2 = f^2h - 2f^2 = -2f^2, hf^3 = -3f^3$, and so on, so that the character of the above representation is precisely $$\mathrm{char}\,\,\Sym\mf{n}^- = 1 + t^{-1} + t^{-2} + t^{-3} + \cdots,$$ in agreement. 

\medskip
We comment on why one may have expected precisely this representation-theoretic prediction from the logic of Harvey-Moore's BPS algebras \cite{HarveyMoore,HarveyMooretwo}. Indeed, Harvey-Moore identify a certain subsector of the BPS algebra with the Generalized Kac-Moody algebra itself, so that one would expect a BPS spectrum organizing into roughly the adjoint representation of the algebra acting on itself. Of course, the BPS algebra is meant to be an honest algebra in this context, not a Lie algebra, and so one should instead account for a second-quantized version thereof, where particle scattering allows for the formation of bound states. We make a final comment on this interpretation of an inverse Weyl denominator BPS spectrum as being organized in a second-quantized Fock space. Indeed, this second-quantization corresponds precisely to why the last step of computations of the BPS spectrum is to take a multiplicative lift of DMVV type, as per section 3 in this paper.

\medskip
We end with some technical remarks: in this and the previous section, we gave our reasoning in the analogous but notably simpler situation of a finite-dimensional Lie algebra, rather than discussing the setting of Generalized Kac-Moody algebras (which actually arise as BPS algebras). We no longer have statements as simple as the definition of a Weyl vector as half the sum of positive roots.  Instead, the defining property of a Weyl vector $\rho$ is now by its pairing against simple roots $\alpha$ as $$\langle \rho, \alpha \rangle = -\frac{1}{2} \langle \alpha, \alpha \rangle$$ and it is not yet proven in all cases of interest that a Weyl vector necessarily exists, although the Fake Monster Lie algebra of interest in this paper does indeed admit a Weyl vector. The fundamental formula above, the Weyl character formula, also changes notably for a Generalized Kac-Moody algebra as root-spaces now admit various mutiplicities and the imaginary simple roots demand their own correction terms.

\section{The Fake Monster Lie algebra}

Combining the above sections, we now argue for the Fake Monster Lie algebra as a symmetry of the heterotic string on $T^7 \times T^2$ (a result partially anticipated by remark 3 on page 28 of \cite{Moore}).
Borcherds gives an expository discussion of his construction of the Fake Monster in \cite{BorcherdsExp}.  Here, we are content to comment that it is the Generalized Kac-Moody one obtains upon suitably enlarging, via the imaginary simple roots, the Kac-Moody algebra with Dynkin diagram the Leech lattice. Notably, this rank $26$ Generalized Kac-Moody algebra has Weyl denominator given by the Borcherds automorphic form $\Phi_{12}$, and so per the BPS count of section $3$ as the inverse of this form, we may now claim that the $1/2$-BPS instantons of heterotic on $T^7 \times T^2$ organize into a second-quantized Fock on the negative part of the Fake Monster Lie algebra. 

\medskip
Note that despite organizing into the named representation $\Sym \mf{n}^-$, the BPS spectrum still changes notably as we vary in moduli space. Most obviously, as we tend towards different Niemeier cusps, we have different numbers of massless states corresponding to the enhanced gauge symmetry arising at the cusp in question; from the current point of view, this enhanced gauge symmetry is the small finite-dimensional part of the Fake Monster Lie algebra that survives unbroken at the specified point in moduli space. More generally, at any given set of charges, the BPS mass (or more suitably, instanton action) varies as we move in moduli space.  This variation is due to the fact that while the Fake Monster acts as a symmetry, it is badly broken, and the generators of the symmetry gain masses that vary in moduli. A more careful investigation of this variation would certainly be of interest.

\medskip
As with any broken symmetry, it is tempting to speculate that at sufficiently high energies or temperatures, the implications of the symmetry nevertheless become
(more) manifest.
It would be very interesting to see if, in a suitable $\alpha^\prime \to \infty$ limit, one could -- for instance -- constrain string scattering amplitudes using identities
generated by the (broken) symmetry.  This would entail investigating the implications of GKM symmetries in the same spirit that higher-spin symmetries have been
intensely investigated in recent years.

\section{Conclusion}

The burden of this paper has been to illustrate, in a particular example, the emergence of a Generalized Kac-Moody symmetry in a concrete string compactification.  In fact the system studied here is one of canonical interest, the simplest compactification of the heterotic string.  The two important considerations we highlighted here are:

\medskip
\noindent
$\bullet$ Fully flavoring BPS counting functions can suggest higher rank GKMs as governing the algebra of BPS states.  While the Igusa cusp form $\Phi_{10}$ arises as the denominator of a rank three algebra in type II strings on $K3 \times T^2$ or (by duality) heterotic strings on $T^6$, upon further compactification and full flavoring, we obtain an algebra of rank 26 in a system with the same supersymmetry.

\medskip
\noindent
$\bullet$ The inverse of the denominator formula giving the automorphic form which counts BPS objects, is a character of a very definite representation of the GKM.  Moreover, this representation is both physically and mathematically natural, and may be said to further explain why the many inverse denominator formulas previously found in the literature are indeed a consequence of the logic of Harvey and Moore.

\medskip
A caveat is that our work has involved compactification to $0+1$ dimensions.  Because of large infrared fluctuations in low dimensions (codified in, for instance, the Coleman-Mermin-Wagner theorem), one does not obtain superselection sectors for different values of the moduli below 2+1 dimensions.  Instead, one obtains a wavefunction on the moduli space.  However, in a limit where the $T^2$ in our $T^7 \times T^2$ has large volume, one should be justified in using the 1-loop computation in three dimensions \cite{Boristwo} as a term which governs the (approximately) three-dimensional physics.  The ``lift'' to $\Phi_{12}$ then encodes the dependence
on the moduli of the (large) $T^2$.

\medskip
There are several directions for further work.  Similar algebraic structures can readily be seen to arise in toroidal compactification of the type II string, and in the
simplest classes of CHL models with half-maximal supersymmetry \cite{toappear}.  There are also relations to enumerative geometry; fully-flavored BPS counting
functions can be given interpretations as flavored curve counts where one takes steps towards fully categorifying the space of BPS states, instead of just computing the coarsest index.  Results such as those in \cite{Lerche}  will readily admit reinterpretation in this light, as providing evidence both for existence of higher rank GKM symmetries, and as illustrations of flavored curve counts in algebraic geometry \cite{toappearalso}.  Finally, it might be interesting to do explicit computations of the algebraic structure governing perturbative sub-sectors of this model (generalizing discussions in e.g. \S3 of \cite{HarveyMooretwo}) to derive subalgebras of the structure studied here.

\bigskip

\centerline{\bf{Acknowledgements}}

\medskip
We would like to thank N. Paquette, R. Volpato, and M. Zimet for very helpful discussions, and J. Harvey and N. Paquette for comments on a draft.   The research of S.K. was supported in part by the National Science 
Foundation under grant NSF-PHY-1316699.

\section{Appendix: A fermionic Fock space construction}

In this appendix, we explain the relation of our representation-theoretic constructions to related constructions of~\cite{Kostant, Kostanttwo, PPV1}, and particularly section $4$ of~\cite{PPV2}. We have the same caveats as in the main text, namely that we will explain the relevant constructions only in the analogous but simpler case of finite-dimensional Lie algebras. Note, however that~\cite{PPV2} manages to explicitly construct the necessary representation in the full GKM case for the monstrous algebras of interest in their work. 

\medskip
Our goal here is to explain that the representation $V_{\rho}$ with highest weight the Weyl vector $\rho$ has a ``fermionic'' character given by the Weyl denominator $\Phi$. Moreover, $$\bigwedge \mf{n}^- \simeq V_{\rho},$$ in that the left-hand side admits a natural irreducible $\mf{g}$-module structure with highest weight $\rho$. By the usual inversion between characters of bosonic and fermionic Fock spaces, this provides a natural link between the bosonic construction we presented here and the fermionic construction of~\cite{PPV2}. We now explicate this fermionic construction.

\medskip
To begin, we compute the (usual) character of $V_{\rho}$. Recall that the Weyl character formula for $\lambda = 0$ already yields an interesting identity for the Weyl denominator, namely $$\sum_{w \in W} \varepsilon(w) e^{w(\rho)} = e^{\rho} \prod_{\alpha \in \Delta^+} (1 - e^{-\alpha}).$$ Doubling all exponents preserves the truth of this formula, and so $$\sum_{w \in W} \varepsilon(w) e^{2 w(\rho)} = e^{2 \rho} \prod_{\alpha \in \Delta^+} (1 - e^{-2 \alpha}) = \Big( e^{\rho} \prod_{\alpha \in \Delta^+} (1 - e^{-\alpha}) \Big) \Big( e^{\rho} \prod_{\alpha \in \Delta^+} (1 + e^{-\alpha}) \Big),$$ implying $$\mathrm{char}(V_{\rho}) = \frac{ \sum_{w \in W} \varepsilon(w) e^{w(2 \rho)}}{e^{\rho} \prod_{\alpha \in \Delta^+} (1 - e^{-\alpha})} = e^{\rho} \prod_{\alpha \in \Delta^+} (1 + e^{-\alpha}).$$ Note that the character of $V_{\rho}$ greatly resembles the Weyl denominator. Of course, the Weyl denominator cannot literally be the character of a (virtual) representation, being Weyl anti-invariant rather than invariant, but if $V_{\rho}$ had some natural grading by a fermion number $F$ and we inserted $(-1)^F$ as a convention into our expression for the character, we could easily arrive at the precise Weyl denominator formula. This step is immediately effected by the identification $V_{\rho} \simeq \bigwedge \mf{n}^-$ for the exterior algebra grading on the right-hand side, so we now turn to explaining this statement.

\medskip
We now proceed to explain why one should think of the total exterior algebra $\bigwedge \mf{n}^-$ as endowed with the structure of a $\mf{g}$-representation. Indeed, we describe how the maximal torus should act: writing out explicitly $$\mf{n}^- = \bigoplus_{\alpha \in \Delta^+} \mf{g}^{-\alpha},$$ we mandate that $\mf{h}$ act on the summand $\mf{g}_{-\alpha_1} \wedge \cdots \wedge \mf{g}_{-\alpha_n} \subset \bigwedge \mf{n}^-$ by the character $$\rho - \sum_{i=1}^n \alpha_i.$$ We hence immediately calculate the character of this $\mf{h}$-representation to be $$e^{\rho} \prod_{\alpha \in \Delta^+} (1 + e^{-\alpha}).$$ We noted above that this expression is $W$-invariant, and hence the character of a (virtual) $\mf{g}$-representation.  Here we see it is,  moreover, explicitly the character of $V_{\rho}$ itself. So, we see that computing a fermionic character with respect to the grading on the exterior algebra, indeed returns the Weyl denominator formula.

\end{document}